\documentclass[aps,prl,twocolumn,showpacs,superscriptaddress,groupedaddress]{revtex4-1}  
\usepackage{graphicx}  
\usepackage{dcolumn}   
\usepackage{bm}        
\usepackage{amssymb}   
\usepackage{subfigure}
\usepackage{natbib}
\usepackage{amsmath}
\usepackage[colorlinks]{hyperref}

\begin{document}

\widetext
\leftline{Version 2.19 as of \today}


\title{Anti-Lenz Supercurrents in Superconducting Spin Valves}

\date{\today}

\begin{abstract}
Here, we present a study on Si(111)/\-Ta($150$\AA )/\-IrMn($150$\AA )/\-NiFe($50$\AA )/\-Nb($x$)/\-NiFe($50$\AA )/\-Ta($50$\AA ) and Si(111)/\-Ta($150$\AA )/\-NiFe($50$\AA )/\-Nb($x$)/\-NiFe($50$\AA )/\-IrMn($150$\AA )/\-Ta($50$\AA ) spin-valves with $x=100$ to $500$\AA . For both sample families, above a specific critical thickness of the Nb-layer and below $T_c$, the superconducting Nb-layer contributes strongly to the magnetization. These systems show an anomalous hysteresis loop in the magnetization of the superconducting layer; the hysteresis loop is similar to what is generally expected from hard superconductors and many superconductor/ferromagnet hybrid systems, but the direction of the hysteresis loop is inverted when compared to what is generally observed (paramagnetic for up sweeping fields and diamagnetic for down sweeping fields). This means that the respective samples exhibit a magnetization, which is contrary to what should be expected from the Lenz' rule.
\end{abstract}

\author{U. D. Chac\'on Hern\'andez}
\affiliation{Centro Brasileiro de Pesquisas F\'isicas, Rua Dr. Xavier Sigaud 150, 22290-180 RJ, Brazil}
\author{M. A. Sousa}
\affiliation{Universidade de Bras\'ilia, Instituto de F\'isica, Campus Universitário Darcy Ribeiro, 70919-970 DF, Brazil}
\author{M. B. Fontes}
\author{E. Baggio-Saitovitch}
\affiliation{Centro Brasileiro de Pesquisas F\'isicas, Rua Dr. Xavier Sigaud 150, 22290-180 RJ, Brazil}
\author{C. Enderlein}
\affiliation{Campus de Duque de Caxias, Universidade Federal do Rio de Janeiro, Estrada de Xerém 27, 25245-390 RJ, Brazil}
\maketitle

\section{Introduction}

Since the discovery of the giant magnetoresistance in 1988 \cite{baibich1988giant, grunberg1988magnetfeldsensor}, spin valves and similar interlayer systems have become a heavily researched topic \cite{Wolf2006}. More recently, spin valves with a superconducting (SC) interlayer have attracted specific attention \cite{potenza, miao2008infinite, Fominov2010, zdravkov2013, patino2013, Banerjee2014, alidoust2014meissner, Leskin2015, linder2015superconducting}. This is closely related to the occurrence of triplet superconductivity in certain ferromagnet(FM)-SC multilayer systems, where spin mixing is present \cite{bergeret2001long, kadigrobov2001quantum, Bergeret2005, Banerjee2014, zdravkov2013, Fominov2010, alidoust2014meissner, Leskin2015, linder2015superconducting}. Moreover, a number of interesting features have been observed in SC spin valves. These include the spin switch- and inverse spin switch effect \cite{gu, potenza, moraru, rusanov2006inverse, zhu2009origin}, superconducting magnetoresistance \cite{Steiner, miao2008infinite, stamopoulos2014absolute, stamopoulos2015superconducting}, and a number of peculiar observations in the magnetization characteristics of the SC layer \cite{stamopoulos2009stray, patino2013}.

Here, we present a study on the magnetization of Si(111)/\-Ta($150$\AA )/\-IrMn($150$\AA )/\-NiFe($50$\AA )/\-Nb($x$)/\-NiFe($50$\AA )/\-Ta($50$\AA ) (denoted as the \textit{bottom} family) and Si(111)/\-Ta($150$\AA )/\-NiFe($50$\AA )/\-Nb($x$)/\-NiFe($50$\AA )/\-IrMn($150$\AA )/\-Ta($50$\AA ) (denoted as the \textit{top} family) devices with $x=100$ to $500$\AA . Samples below certain critical Nb-layer thicknesses ($300$\AA\ for \textit{bottom} and $350$\AA\ for \textit{top}) exhibit a magnetization curve as it is commonly found for spin valves. Samples with higher Nb-layer thicknesses exhibit a strong hysteresis loop, similar to what is known from hard SCs and a number of interlayer compounds. However, in strong contrast to such hysteresis loops, the present ones are in the opposite direction, as one would expect. Thus, the respective samples are naturally paramagnetic in the SC phase (for up-sweeping external fields), and become diamagnetic for down-sweeping fields. Therefore, the magnetization behaves contrary to what one would expect from Lenz' rule, which is generally used to explain the magnetization behavior of hard SCs. Thus, the respective samples always generate a field directed in such manner that it amplifies the change of an external field. 

The anomalous hysteresis loop (clockwise in the M-H diagram) is the main focus of this paper. Thus, we will often refer to systems with a \textit{normal} hysteresis loop (counter-clockwise in the M-H diagram). For simplification, we will always refer to such systems as \textit{hard SCs}, despite the fact that cylindrical superconductors, many FM/SC interlayer systems and a variety of other samples exhibit the same hysteresis loop, without being what is generally termed \textit{hard SC}.

Moreover, we will refer to samples by using the thickness of the Nb layer and the family of the respective sample. Thus, when referring to \textit{the 400\AA\ top sample}, we speak of the sample of the \textit{top} family with a 400\AA\ thick Nb layer.

\section{Experiment}

\subsection{Sample Preparation}

\begin{table*}[htbp]
	\centering
		\begin{tabular}{|cccccccc|} \hline 
    Sistema &  $d_{FM1}$ (\AA) & $h_{FM1}$ (\AA) & $d_{SC}$ (\AA) & $h_{SC}$ & $d_{FM2}$ (\AA) & $h_{FM2}$ (\AA) & $\chi_i^{2}$ \\ \hline\hline
      NiFe/Nb(100 \AA)/NiFe \textit{bottom} & 50.2 & 6.4   & 109.0     & 4.3  & 51.6  & 8.8 & 0.61 \\
      NiFe/Nb(100 \AA)/NiFe \textit{top} & 51,3 & 7.0    & 103.4  & 9.9  & 52.3  & 9.3  & 0.63 \\
      NiFe/Nb(150 \AA)/NiFe \textit{bottom} & 55.7 & 5.6   & 153.9   & 3.8  & 50.4  & 2.6 & 0.75 \\
      NiFe/Nb(150 \AA)/NiFe \textit{top} & 53.5 & 4.6  & 156.3  & 7.2  & 57.0    & 5.9  & 0.48 \\
      NiFe/Nb(200 \AA)/NiFe \textit{bottom} & 52.5 & 3.6   & 195.3   & 4    & 53.3  & 5.7 & 0.99 \\
      NiFe/Nb(200 \AA)/NiFe \textit{top} & 52.7 & 5.3  & 212.3  & 6.5  & 50.8  & 5.1  & 0.72 \\
      NiFe/Nb(250 \AA)/NiFe \textit{bottom}& 50.4 & 4.2   & 247.2   & 4.5  & 49.9  & 4.5 & 0.62 \\
      NiFe/Nb(250 \AA)/NiFe \textit{top} & 51.4 & 5.3  & 249.9  & 6,2  & 53.3  & 5.5  & 0.52 \\
      NiFe/Nb(300 \AA)/NiFe \textit{bottom}& 51.4 & 3     & 300.5   & 3.2  & 52.9  & 3.2 & 0.65 \\
      NiFe/Nb(300 \AA)/NiFe \textit{top} & 52,4 & 3,1  & 300.4  & 5.3  & 54.9  & 4.8  & 0.62 \\
      NiFe/Nb(350 \AA)/NiFe \textit{bottom}& 52.4 & 2.6   & 352.4   & 3.5  & 50.9  & 3.5 & 0.62 \\
      NiFe/Nb(350 \AA)/NiFe \textit{top} & 57.4 & 2.4 & 351.0    & 2.2   & 56.9 & 2.2 & 0.55 \\
      NiFe/Nb(400 \AA)/NiFe \textit{bottom}& 52.5 & 3.0     & 411.0     & 2.2  & 52.4  & 2.4 & 0.63 \\ 
      NiFe/Nb(400 \AA)/NiFe \textit{top} & 53.2 & 2.6  & 403.5  & 4.4  & 54.4  & 3.0 & 0.84 \\
      NiFe/Nb(500 \AA)/NiFe \textit{bottom}& 52.3 & 1.0     & 500.3   & 2.1  & 51.3  & 1.1 & 0.46 \\ \hline
		\end{tabular}
	\caption{ \small Thickness ($d$) and roughness ($h$) of the respective layers FM1, SC, FM2 as calculated by WINGIXA from X-ray reflectivity. FM1 refers to the layer between the AFM and the SC. $\chi_i^{2}$ designates the respective error, as calculated by WINGIXA.}
\label{table:thicknesses}
\end{table*}

Our samples have been prepared by magnetron sputtering at room temperature on naturally oxidized Si(111) single-crystalline substrates from EPITECH. Before installing the Si substrate on the sample holder in the vacuum chamber, the surface of the crystals has been carefully cleaned with ph-neutral soap, distilled water, and acetone. The base pressure of the chamber before preparation was $4.0 \times 10^{-8}$ Torr and the deposition was performed in an Ar atmosphere with a pressure of 3 mTorr. We used commercial Ta, Ni$_{81}$Fe$_{19}$, Ir$_{20}$Mn$_{80}$,and Nb targets with 99.99\% purity from AJA international for the layer growth. The deposition rates of the targets  have been defined by X-ray reflectivity measurements on films of single composition prepared as calibration samples. As a first step, we chose to prepare a 150\AA\ Ta buffer layer on top of the substrate to favor the smooth growth of the next layers. The same target was used to deposit a 50 \AA\ thick capping layer in order to prevent oxidization. All spin valves were grown under a static magnetic field of 400 Oe applied in the plane of the substrate to induce a unidirectional anisotropy. 

We prepared two different sample families: those samples with  Ta(Buffer)/IrMn/NiFe/Nb/NiFe/Ta(Capping) we call \textit{bottom} and a second family with Ta(Buffer)/NiFe/Nb/NiFe/IrMn/Ta(Capping) we call \textit{top}. The respective thickness and roughness values obtained by WINGIXA  are shown in Table \ref{table:thicknesses}. The thickness and roughness values designated by \textit{FM1} correspond to the layer in contact with the antiferromagnetic (AFM) layer (pinning layer). Those designated by \textit{FM2} refer to the free FM layer.

\subsection{Measurements}

\begin{figure}[ht!]
\centering
\includegraphics[scale=0.29]{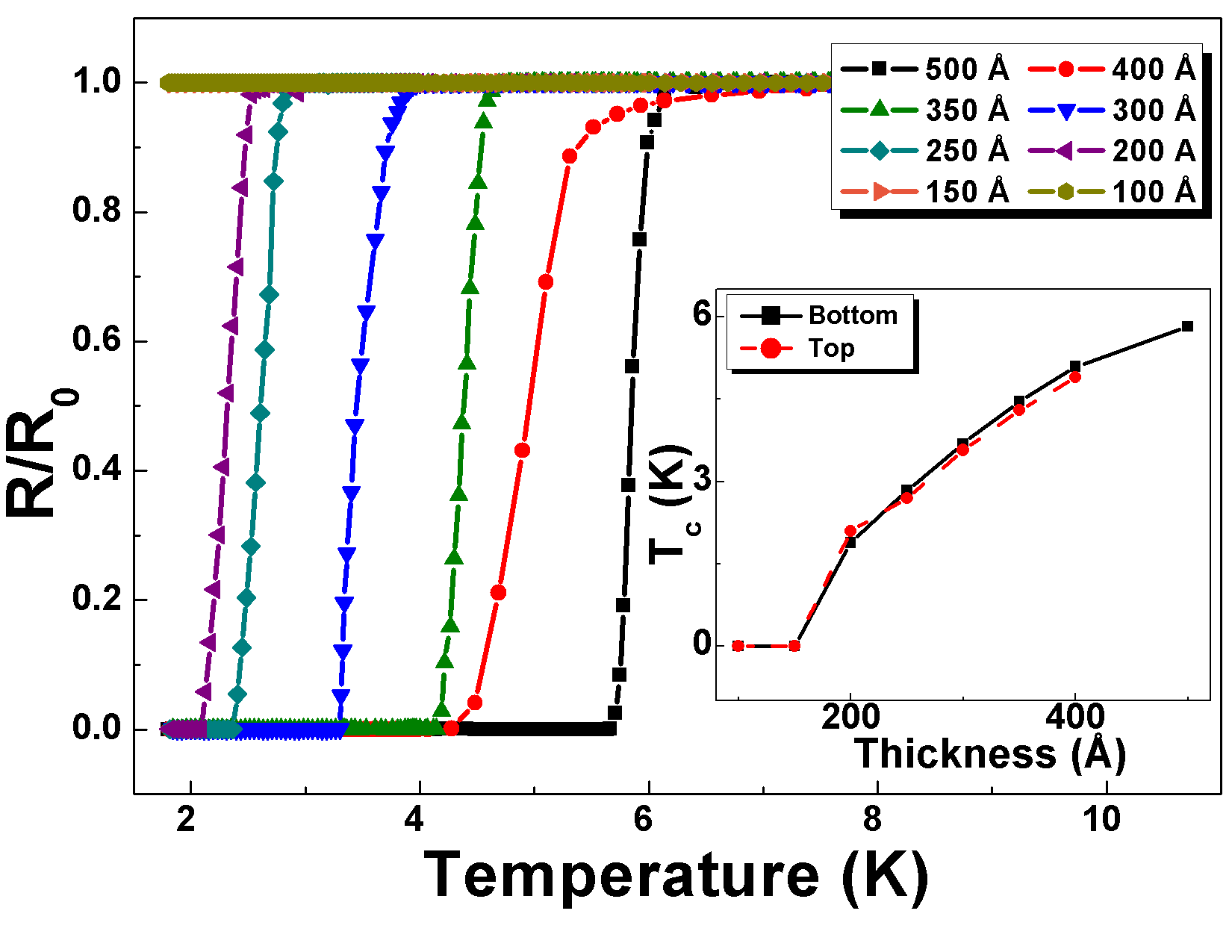}
\caption{\label{fig:res} Resistivity curves for all \textit{bottom} samples. The inset shows the Nb-layer thickness dependence of the critical temperature for all samples.  $T_c$ has been determined by building the derivative of the resistivity curves and defining the maximum as $T_c$.}
\end{figure}

All samples have been studied after preparation by x-ray reflectivity to determine the exact thickness of the layers and the roughness of the interfaces. The characterization has been performed by using a PANALYTICAL X PERT spectrometer and subsequent  use of WINGIXA as the tool of analysis. Resistivity measurements were performed by the four-point method using a LR-700 bridge and a cryomech dry cryostat with $2$K minimum temperature.  Magnetization measurements were performed using the VSM option of a commercial Quantum Design cryofree Dynacool PPMS. To observe the temperature dependence of the magnetization loop, field sweeps have been performed at fixed temperatures between $2$ and $10$K.

\begin{table}[htbp]
	\centering
		\begin{tabular}{|ccc|} \hline 
    Layer &  $d_x$ (\AA) & $h_x$(\AA)  \\ \hline\hline
			Ta (\textit{Buffer})    & $150.2 \pm 0.7$  &  $6.5 \pm 0.7$ \\
      IrMn (AFM)              & $157.3 \pm 0.7$  &  $7.6 \pm 0.7$ \\ 
			Ta (\textit{capping})   & $52.6  \pm 0.7$  &  $2.5 \pm 0.7$ \\ \hline
		\end{tabular}
	\caption{\small Thicknesses of the buffer-, protection-, and antiferromagnetic layers. The nominal error is from averaging over all samples.}
	\label{tab3-Py/Nb/Py}
\end{table}

\begin{figure*}[ht!]
\centering
\includegraphics[scale=0.65]{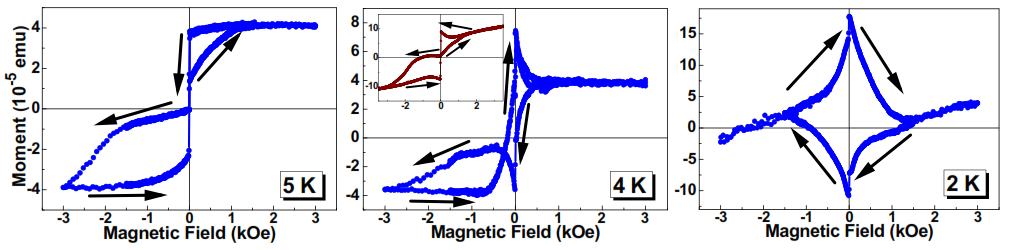}
\caption{\label{fig:Curve} Representative magnetization curves of the \textit{top} sample with a Nb layer thickness of $400$\AA . The critical temperature, as determined by resistivity measurement, accounts for $4.89$ K for this sample. The arrows mark the hysteresis direction. The inset shows the 2K curve of the 350 \AA\ \textit{top} sample.}
\end{figure*}

\section{Results}

The measured sample thicknesses are shown in Table \ref{table:thicknesses} for the ferromagnetic and Nb-layers, while the thicknesses of the other layers are summarized in Table \ref{tab3-Py/Nb/Py}. 

The critical temperatures were determined by resistivity measurements. The respective resistivity curves are shown in Fig. \ref{fig:res}.

Representative magnetization curves are shown in Fig. \ref{fig:Curve}, where one can see magnetization loops for the $400$\AA\ \textit{top} sample for three different temperatures. Fig. \ref{fig:Curve} gives a qualitative understanding of the development of the hysteresis loop with temperature. The critical temperature of this sample, as determined by resistivity measurement, is $4.89 $ K. Thus, the magnetization curve at $5$~K shows the typical behavior of a normal spin valve with a non-SC metallic interlayer. Such a magnetization curve is also found for our thinner superconducting samples of similar design (data not shown in this paper), and this behavior is consistent with previous studies from other groups \cite{Banerjee2014, moraru}. However, for samples with thicker Nb-layers, we see a strong change in behavior, as is demonstrated in the $4$K curve in Fig. \ref{fig:Curve}. One can see clearly that the superconductor magnetizes in a manner, which is opposite to what is generally expected: for up-sweeping fields, the Nb-layer amplifies the external field instead of expelling it. For down-sweeping fields the superconductor expels the field. This effect is strongly enhanced at $2$K. The respective diamond-shaped curve is known from hard type 2 superconductors, however, in our case, the hysteresis loop goes in the opposite direction. 

\begin{figure*}[ht!]
\centering
\includegraphics[scale=4.4]{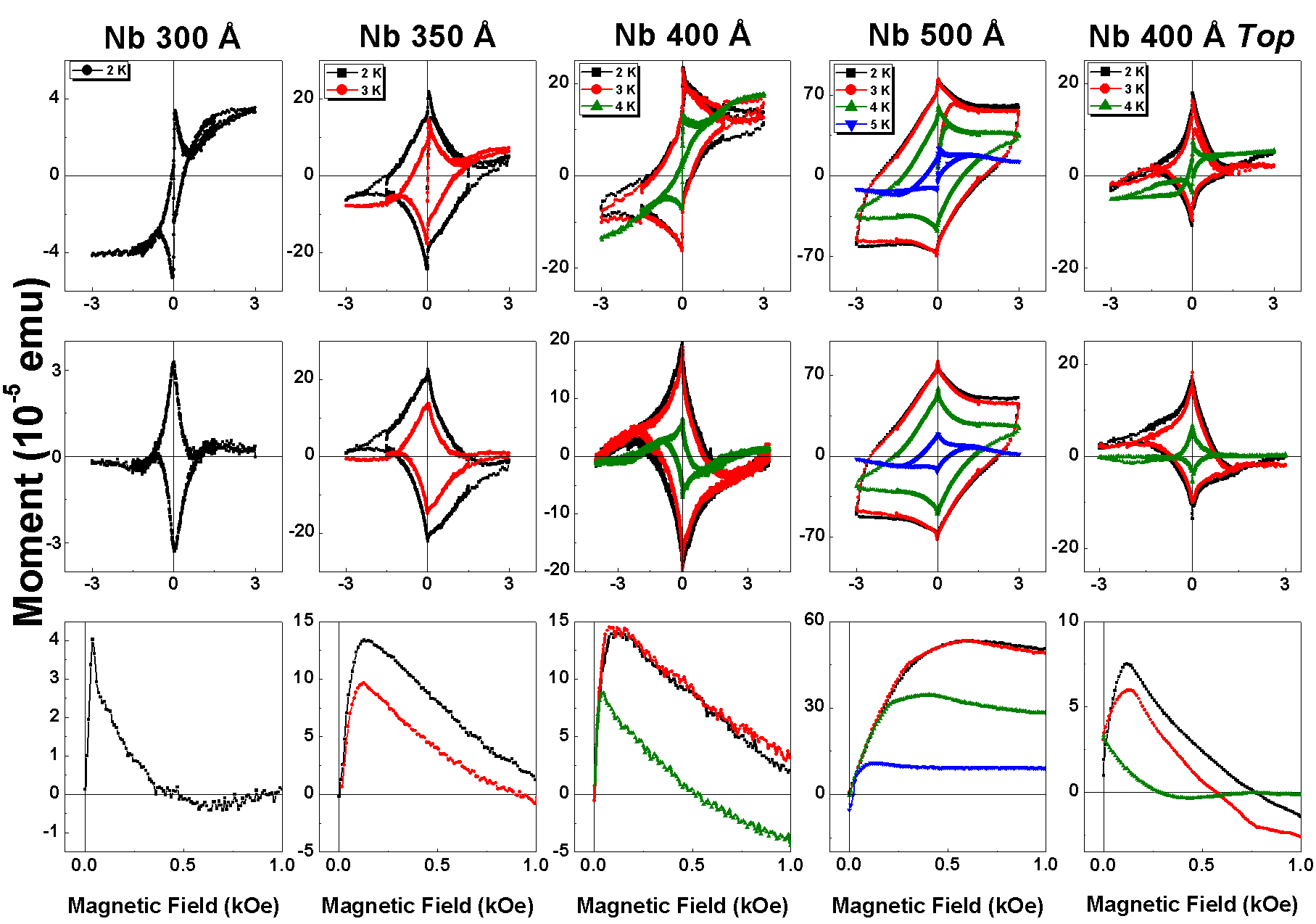}
\caption{\label{fig:sus} First row: Original hysteresis loops for all samples, which show the uncommon hysteresis loop. Second row: Same curves with the subtracted ferromagnetic contribution (by subtracting the magnetization curve slightly above $T_c$). Third row: Initial magnetization curve from zero external field. Samples which are not explicitly denoted as \textit{bottom} are of the \textit{top} type.}
\end{figure*}

Almost all of the measured samples with Nb-layer thicknesses above 300\AA\ exhibit the reversed hysteresis loop. The only exception is given by the 350\AA\ \textit{top} sample, for which the 2K curve is shown in the inset of Fig. \ref{fig:Curve}. As one can see in Table \ref{table:thicknesses} , this specific sample possesses about $10$\%\ thicker NiFe layers.

To better visualize the magnetization of the superconducting layer, we subtracted the ferromagnetic contribution from the original magnetization curves. For each sample, the magnetization curve for the lowest temperature above $T_c$ has been subtracted from the measured curves below $T_c$. The result is shown in the middle row of Fig. \ref{fig:sus}.

Naturally, the subtraction of the ferromagnetic curves leads to a minor glitch in the data around zero external field, due to a change of exchange bias. Thus, the small thin peak at the respective position can be ignored. 

An important feature of the hysteresis loop of a hard SC is the so-called intermediate reversible state, which can be found at the outer ends of the hysteresis loop, where the samples are still superconducting but the hysteresis is closed. It is interersting that the intermediate phase is not the same for all samples, which show the clockwise hysteresis loop. In case of the \textit{bottom} samples, the $300$\AA\ sample exhibits a paramagnetic maximum in the magnetization at around $150$ Oe, while at the closing point of the hysteresis, the magnetization is almost neutral. The $350$\AA\ \textit{bottom} sample exhibits diamagnetic behavior in the respective phase, as in normal hard superconductors.  This seems to be equally true for the $400$\AA\ \textit{bottom} sample, while the $500$\AA\ sample is again paramagnetic in the reversible state. In contrast to this, the $400$\AA\ \textit{top} sample exhibits a diamagnetic reversible state.

In the third row of Fig.\ref{fig:sus}, we see the initial curves from zero magnetization for all samples. As is obvious, all shown curves exhibit a very strong paramagnetic behavior. The \textit{bottom} samples with Nb-layer thicknesses between $300$ and $400$\AA\ exhibit a magnetization maximum below $250$Oe for all temperatures. Then the magnetization drops and crosses zero for higher fields. The same behavior is observed for the $400$\AA\ \textit{top} sample. The $500$\AA\ \textit{bottom} sample shows, in contrast to this, a plateau after reaching a certain maximum magnetization.

All of the discussed samples show the paramagnetic response from zero magnetization and have the full clock-wise hysteresis loop in common. This means that the superconducting samples act first paramagnetic, which is more commonly observed under field cooling in a variety of superconducting samples (the connection to the paramagnetic Meissner effect will be discussed further below). However, in our samples the fully closed hysteresis loop suggests that the paramagnetism, or - more general - the anti-Lenz law behavior, is intrinsic in nature.

\section{Discussion}

The observed effect is particularly novel and seemingly inconsistent with common believes on superconductors. Therefore, one should be sure of absence of any artifact or the result of faulty measurement procedures. 

There are a number of reasons, why a VSM measurement, as it has been used for this study, can actually produce inversed magnetization curves. Obviously, a trivial response from a hard superconductor, when inverted, would partly show the observed behavior. However, if the magnetization direction was inverted, we would not see the commonly expected correct magnetization in the non-superconducting phase. 

Moreover, as mentioned, we have one sample which shows the classical behavior (the $350$\AA\ \textit{top} sample as shown in the inset of Fig. \ref{fig:Curve}), which possesses $10$\%\ thicker ferromagnetic layers. 

Finally, one can see in the $4$ K curve in Fig. \ref{fig:Curve} and in several curves in Fig. \ref{fig:sus}, the crossing of the up-sweeping and the down-sweeping curves. This has been found in all measured samples. Thus, we can exclude the possibility of the switch of magnetization just being an artefact from the magnetization measurement method, since in the intervals of temperature, where the sample is not superconducting, reproduce what is well known from similar samples. 

\subsection{Vortex- or supercurrent structure}

The unusual phenomenon of intrinsic paramagnetism in a superconductor stands in conflict with common believes concerning the superconducting state. We therefore should be rather careful about statements of the nature of the magnetization. Specifically, one can wonder, if the paramagnetism originates from vortices, which penetrate an otherwise diamagnetic sample. If this was the case, the amount of vortices would be reduced for up-sweeping fields, contrary to what is known from hard superconductors. One could also consider that the magnetization is a consequence of macroscopic supercurrents, which either become weakend, for example, by a decreasing order parameter, or which are penetrated by diamagnetic vortex-like structures.

\begin{figure}[ht!]
\centering
\includegraphics[scale=2]{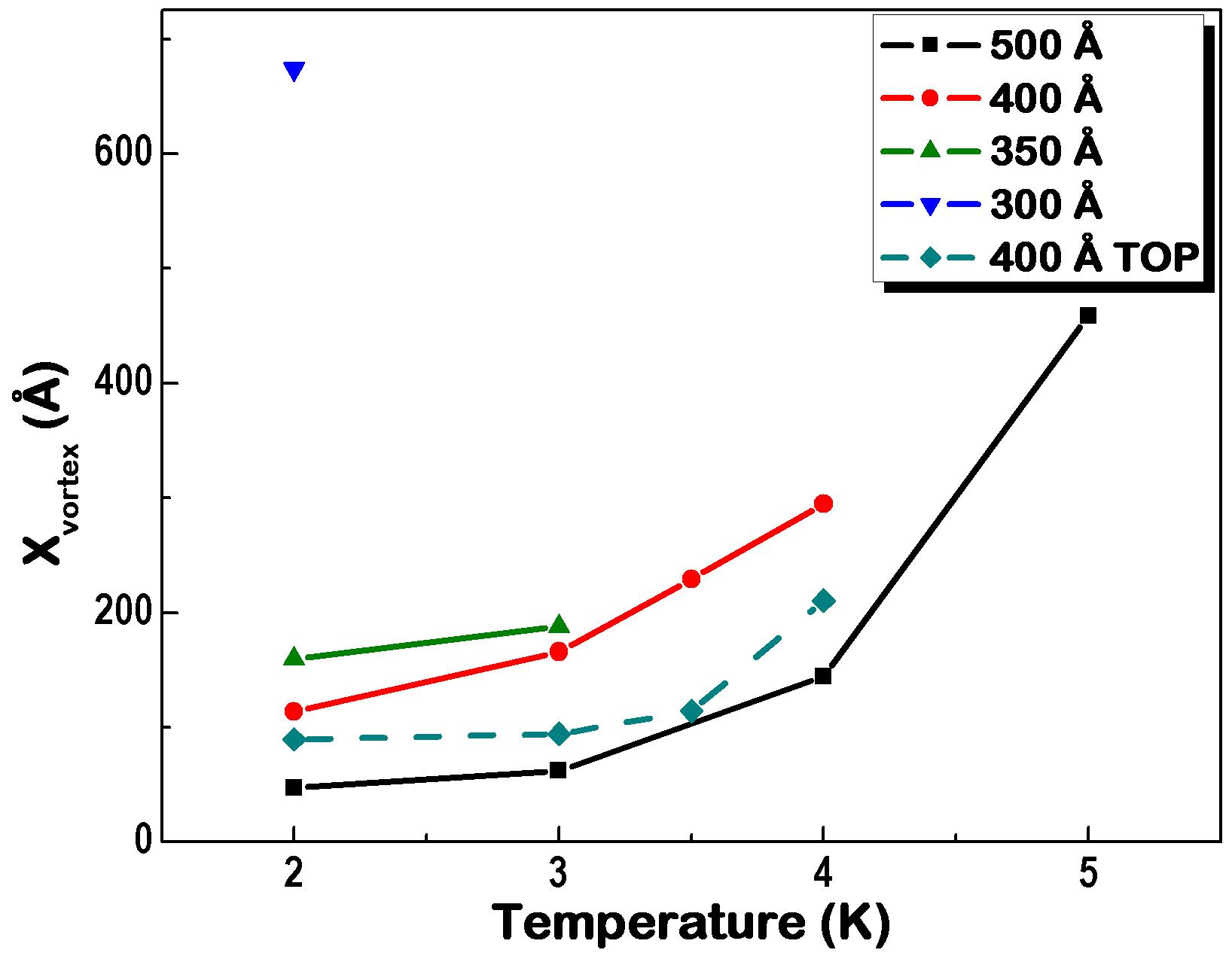}
\caption{\label{fig:vor} Speculative distance between vortex cores, considering that each vortex core only carries on quantum of flux.}
\end{figure}

To answer this question, we first assume that the field is generated through singular vortices, each carrying one quantum of flux. Since we know the sample dimensions and the magnetization in zero external field, we can make an estimate for the distance between the vortex cores in the $x$ direction (in-plane direction perpendicular to the external field direction). The assumption that each vortex carries only one quantum of flux is actually bold, since it has been shown that vortices can carry several quanta of flux for samples which exhibit the paramagnetic Meissner effect, which is the most similar phenomenon studied \cite{Geim1998}. However, it is a meaningful, though speculative, approach to see how far the vortex cores would be from each other in the x-direction. The result of this analysis is shown in Fig.\ref{fig:vor}, indicating that the distance between vortex cores decreases for lower temperatures, as well as for thinner Nb-thickness. At some temperature, the magnetization, and with it the supposed vortex density, seems to saturate. This value below $100$\AA\ is clearly too low to justify the idea that the current is carried through singular vortices. Thus, even if one assumes the vortices heavily penetrating the surrounding layers (where the order parameter actually should be lower, resulting in a larger vortex size), the result suggests that it is unlikely that the magnetic moment is generated by singular vortices.

So, we will consider the other extreme: one singular rectangular supercurrent. In that case, the superconducting order parameter penetrates the ferromagnet and, as is well known, oscillates. We would naturally expect the formation of a triplet compound close to the interfaces. Interestingly, the neighboring Ta is close to a superconducting instability and might actually allow a relatively deep penetration of the superconducting order parameter, which might be somehow related to the observed behavior

\subsection{On the influence of stray fields}

It is tempting to assume that somehow stray fields are involved in the reversed magnetization. Although, it might be trivial for some, we will show in this subsection that an explanation based on this idea can be excluded. 

The stray fields from the magnetizing permalloy layers must naturally be directed in the opposite direction to the external field. However, it is important to note that it can be excluded that the stray field of the magnetizing permalloy layers is responsible for the observed effect. This can be easily shown by calculating the stray field from the permalloy layer. This accounts trivially for

\begin{widetext}
\begin{equation}
\vec{B}(\vec{r}) = \frac{\mu_0}{4\pi}  \int_{V_{\lambda}} \frac{ 3 (\vec{r}-\vec{r}_{\lambda}) (\vec{\lambda} (\vec{r}-\vec{r}_{\lambda})) - \vec{\lambda}(\vec{r}-\vec{r}_{\lambda})^2}{ | \vec{r}-\vec{r}_{\lambda}|^5} dV_{\lambda}
\end{equation}
\end{widetext}

with $\lambda$ being the magnetic moment density and $V_{\lambda}$ being the volume of the layer. Our samples have roughly an area of $3\times 3$mm$^2$ and, thus, one can easily calculate the field by numerically integrating with Mathematica. From our measurements, we know that each permalloy layer has roughly a total magnetic momentum of $6\cdot 10^{-5}$emu and find therefore a $\lambda$ of $1.3\cdot 10^{9}$emu$/$m$^3$. This gives a constant strayfield of $1.2\cdot 10^{-2}$Oe within the Nb layer. Obviously, this field is so small that it is not relevant for any effects.

However, stray fields from magnetic domains are, of course, significantly higher, but since the hysteresis loop is completely reversible, we assume that the measured effect is even present, when the whole permalloy layer is magnetized. This would contradict the idea that domain walls might be responsible for the observed behavior.

\subsection{On the connection to the paramagnetic Meissner effect}

Paramagnetism in superconductors has been observed under field cooling in high temperature superconductors \cite{svedlindh1989anti, braunisch1992paramagnetic, braunisch1993paramagnetic, riedling1994observation, okram1997paramagnetic, chou1993preparation, golovashkin2000anomalous}, in small or porous samples of Nb, Al, and Pb \cite{thompson1995observation, kostic1996paramagnetic, Geim1998, charnaya2013paramagnetic, yuan2004paramagnetic}, and layered systems with superconducting and magnetic materials \cite{lee2003paramagnetic, xing2009controlled, di2015intrinsic}. It has been suggested that the paramagnetic Meissner effect (PME) can arise due to vortex fluctuations combined with pinning \cite{svedlindh1989anti}, phase-shifts of $\pi$ originating from disorder or impurities \cite{bulaevskii1977superconducting, spivak1991negative, kusmartsev1992destruction, sigrist1995unusual}, and intrinsic $\pi$-junctions in $d$-wave superconductors \cite{sigrist1992paramagnetic}. More recently, the paramagnetic response due to superconductors in contact to spin-active interfaces has extensively been discussed \cite{yokoyama2011anomalous, asano2011unconventional, mironov2012vanishing, alidoust2014meissner, di2015intrinsic}. Obviously, what is generally referred to as the PME, is a wide range of phenomena, which can hardly be considered as being equal. However, one can say that what is mostly called PME is an effect which happens under field-cooling and the paramagnetic state is meta-stable. Since our samples tend to behave paramagnetic and only under down-sweeping fields diamagnetic, we must assume that the observed state is a fully stable one, while the diamagnetic state is metastable.

However, there have been experiments, which showed a similar behavior. Stamopoulos et al, in their 2009 arxiv preprint \cite{stamopoulos2009stray}, showed a paramagnetic response of the superconducting layer for transverse fields in Ni$_{80}$Fe$_{20}$-Nb-Ni$_{80}$Fe$_{20}$ trilayers. However, their interpretation in terms of stray fields is in their case very consistent and, thus, the origin of this effect is fundamentally different to what has been observed in our case. A more similar example is by Di Bernardo and collaborators, who were able to detect an intrinsic paramagnetic response of Cooper pairs in the gold layer of Au/Ho/Nb trilayers \cite{di2015intrinsic}. 

\section{Conclusions}

We presented the unusual effect of a clockwise hysteresis loop in the M-H diagram in superconducting spin valves. The effect appears in NiFe($50$\AA )/Nb($x$)/NiFe($50$\AA )/IrMn($150$\AA ) spin-valves with $x=350$ to $500$\AA spin valves and one of the measured samples suggests that it disappears for thicker NiFe layers. Our analysis suggests that the current does not originate from singular vortices. Moreover, although we can exclude the involvement of stray fields.

\begin{acknowledgments}
The authors would like to thank J. Litterst and M.A. Continentino for fruitful discussions. CE would like to thank ICAM and FAPERJ for continuous funding over the period of the project.
\end{acknowledgments}

\end{document}